# Anisotropic permeability in deterministic lateral displacement arrays†

Rohan Vernekar, *[a] Timm Krüger, [a] Kevin Loutherback,‡[b] Keith Morton[c] and David W. Inglis[d]

We uncover anisotropic permeability in microfluidic deterministic lateral displacement (DLD) arrays. A DLD array can achieve high-resolution bimodal size-based separation of microparticles, including bioparticles, such as cells. For an application with a given separation size, correct device operation requires that the flow remains at a fixed angle to the obstacle array. We demonstrate *via* experiments and lattice-Boltzmann simulations that subtle array design features cause anisotropic permeability. Anisotropic permeability indicates the microfluidic array's intrinsic tendency to induce an undesired lateral pressure gradient. This can cause an inclined flow and therefore local changes in the critical separation size. Thus, particle trajectories can become unpredictable and the device useless for the desired separation task. Anisotropy becomes severe for arrays with unequal axial and lateral gaps between obstacle posts and highly asymmetric post shapes. Furthermore, of the two equivalent array layouts employed with the DLD, the rotated-square layout does not display intrinsic anisotropy. We therefore recommend this layout over the easier-to-implement parallelogram layout. We provide additional guidelines for avoiding adverse effects of anisotropy on the DLD.

## 1 Introduction

Deterministic lateral displacement (DLD) is a hydrodynamic size-based particle separation technique. DLD is capable of high-resolution separation of particles up to five times smaller than the array gap ($G$ in Fig. 1A). The DLD method[1] can be used with various types of particles and has shown promise in separation and purification of bioparticles. This technique employs an inclined obstacle array in a microfluidic channel. The array inclination determines the particle separation size (critical radius, $r_c$) in the DLD device. The conceptual framework for understanding and designing DLD arrays is based on the assumption that the locally averaged fluid flow direction remains at a fixed angle to the obstacle array incline throughout the device.

The fixed inclination creates a regular and uniform pattern of fluid flow lanes separated by stagnation streamlines in the microchannel. The width of the flow lane adjacent to every obstacle determines the critical particle radius,[2] $r_c$. For dilute suspensions,[3] particles larger than $r_c$ follow the array inclination, and particles smaller than $r_c$ are advected along the fluid streamlines. In the high Péclet number limit (advection dominating over diffusion), particle paths are deterministic. The path for particles with a radius $<r_c$ is called the "zigzag" trajectory as the smaller particles move laterally back and forth while following the fluid streamlines. For particles of radius $>r_c$, the path is named the "bump" trajectory; these particles are bumped into adjacent streamlines by an obstacle post at every row and follow the array inclination.

The DLD technique has the advantage of being label-free, relying solely on hydrodynamic and volume exclusion forces to achieve separation. This technique has been demonstrated for various applications such as microbead separation,[1,2,4] fractionation of human blood components,[5–9] separation of parasites or circulating tumour cells from human blood[10–12] and deformability-based mapping of human blood.[13–15] Additionally, various array post shapes such as square, circle, triangle (right/equilateral), I-shape, L-shape *etc.* have been employed to obtain improved DLD separation.[16–19]

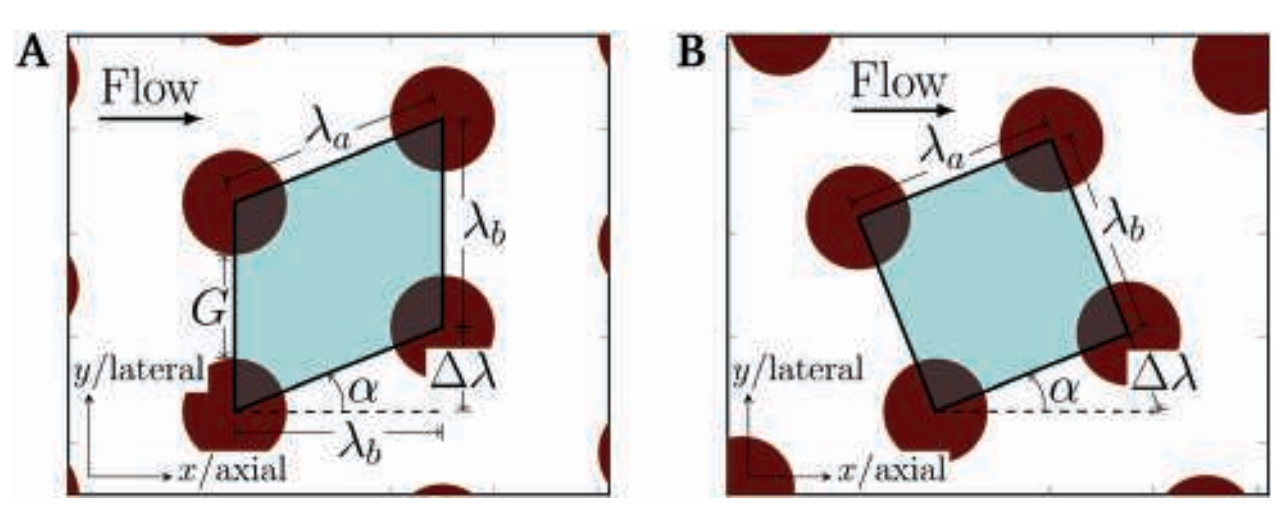

**Fig. 1** (A) Row-shifted parallelogram layout and (B) rotated-square layout employed in DLD arrays. Note the lateral (up–down) and axial (flow-wise) directions.

[a] School of Engineering, University of Edinburgh, King's Buildings, Edinburgh, UK. E-mail: R.Vernekar@ed.ac.uk; Tel: +44 (0)131 6505685
[b] Lawrence Berkeley National Lab, Berkeley, CA, USA
[c] Life Sciences Division, National Research Council of Canada, 75 de Mortagne Boulevard, Boucherville, QC J4B 6Y4, Canada
[d] Department of Engineering, Macquarie University, Sydney, Australia

† Electronic supplementary information (ESI) available. See DOI: 10.1039/c7lc00785j

‡ Current address: Berkeley Lights, Inc., Emeryville, CA, USA.

Across all published works, array posts of any shape are arranged in one of these two layouts: row-shifted parallelogram (Fig. 1A) or rotated-square layout (Fig. 1B). In the parallelogram layout, adjacent rows of posts are shifted by a fixed amount $\Delta\lambda$, which results in a parallelogram unit cell, and the array has an inclination of $\tan^{-1}(\Delta\lambda/\lambda_b)$ with respect to the horizontal flow. In the rotated-square layout, a Cartesian periodic array (unit square cell length $\lambda_a = \lambda_b$) is rotated by the required angle $\alpha = \tan^{-1}(\Delta\lambda/\lambda_b)$. Therefore, the ratio $\Delta\lambda/\lambda_b$ determines the array inclination and is termed row shift fraction $\varepsilon$. When $\varepsilon$ is given as 1 over an integer ($\varepsilon = 1/N$), bimodal particle separation is expected.[21] $N$ gives the periodicity of the array.

Both parallelogram and rotated-square layouts are equally popular for separation applications; several authors[5–8,16,17,20–24] have used parallelogram arrays, and others[1,13,18,19,25,26] have employed the rotated-square layout. Parallelogram arrays are attractive and easier to design because they have a planar boundary (interface) that is perpendicular to the axial flow direction. This is also an advantage when placing arrays with different separation angles in series (cascaded arrays). However, the consequences of the differences between the two layouts have not been understood, and these are used interchangeably.

We show that the parallelogram layout, unlike the rotated-square layout, suffers from certain drawbacks (section 2). These disadvantages include array regions where particle separation does not occur at all or has a different critical size or even a negative separation angle.[21,27,28] This would lead to particles not separating into distinct bands according to the size range and not exiting at the intended outlet ports.

Through extensive lattice-Boltzmann simulations,[29] we show that the issues with the parallelogram layout arise from array-induced anisotropy (sections 3 and 4). Anisotropy is the tendency of an array to induce a lateral pressure drop to the main flow direction in a device. This anisotropic lateral pressure induces a secondary background re-circulatory flow in the presence of certain design features, which we call "enablers". The secondary flow causes local deviations in the primary flow field such that the flow no longer remains parallel to the side walls everywhere in the device. Thus, away from the device side walls, the primary flow tilts and this causes spatially varying critical separation size, which impairs deterministic separation.

We discover that an interface gap (before/after array sections) and large jumps in the array inclinations with cascaded arrays act as enablers (section 4.4). Such features are common in DLD devices and should be avoided. We also find that anisotropy becomes acute when unequal axial and lateral array gaps are employed[30] and when highly asymmetric post shapes are used (section 5). However, we find that the rotated-square layout with circular posts possesses no anisotropy. Array anisotropy can cause significant alterations in expected particle separation trajectories and needs to be accounted for by the design of microfluidic devices that use obstacle arrays[31–33] (section 6).

# 2 Consequences of array anisotropy

In this section, we present experimental evidence of flow tilt and its consequences on particle trajectories in the parallelogram array layout. We also observe that in an equivalent rotated-square layout, the flow does not tilt and remains along the horizontal.

### 2.1 Parallelogram layout

In an experiment intended for particle separation and particle crossing, we observed a significant tilt of the fluid flow away from the horizontal at the interface between two DLD array inclinations. As detailed in Fig. 2, a jet of red fluorescent, 2.7 μm beads mixed with green fluorescein dye was injected into the DLD array between two, co-flowing streams of clear buffer. The 2.7 μm beads immediately follow the standard bump mode, but the fluorescent dye deviates from the horizontal in the region around the interface between positive and negative array inclinations. The positions of these two array sections along the length of the device are shown schematically in Fig. 2A. The DLD design used here is based on a parallelogram-type array layout (Fig. 2F) with 11 μm horizontal and vertical post–pitch. The array inclinations are set to $\alpha = \pm 11.3°$ ($\varepsilon = 1/5$). The cylindrical posts are 7.3 μm in diameter giving a lateral gap between posts of 3.7 μm. The DLD devices were fabricated on polished silicon substrates (Fig. 2B) using standard photolithography techniques and deep reactive ion etching to create vertical sided posts to a depth of 18 μm. Sequential parallelogram array sections are placed directly one after another without any interface structure between opposite inclinations.

Fig. 2D is a series of time exposure images captured with a colour CCD camera and then stitched together to reconstruct the overall motion of the beads and dye from the injection point through to the second array section. The 2.7 μm beads (red) clearly track in the bump mode, following the array inclination. However, the path of the fluorescent dye (green) deviates noticeably from the horizontal. From the zoomed-in image in Fig. 2E, we can see that the fluorescent green dye shows a distinct tilt, preferentially following the prevailing array inclination. This tilt is especially noticeable as the dye advects across the interface between the two sections. This junction between the two arrays is shown in detail in the top view SEM image in Fig. 2F. This experiment captures an anisotropic flow tilt in the parallelogram-type DLD device layout. Here, the trajectory of particles in the bump mode remains unaffected because the particle size of 2.7 μm is larger than the critical particle size for this device of $d_c$ = 2.4 μm. But the anisotropic flow tilt can perturb trajectories of smaller particles travelling in the zigzag mode as shown in Fig. 3.

In a second experiment, the same DLD device design was used, but with an input jet of beads with different diameters.

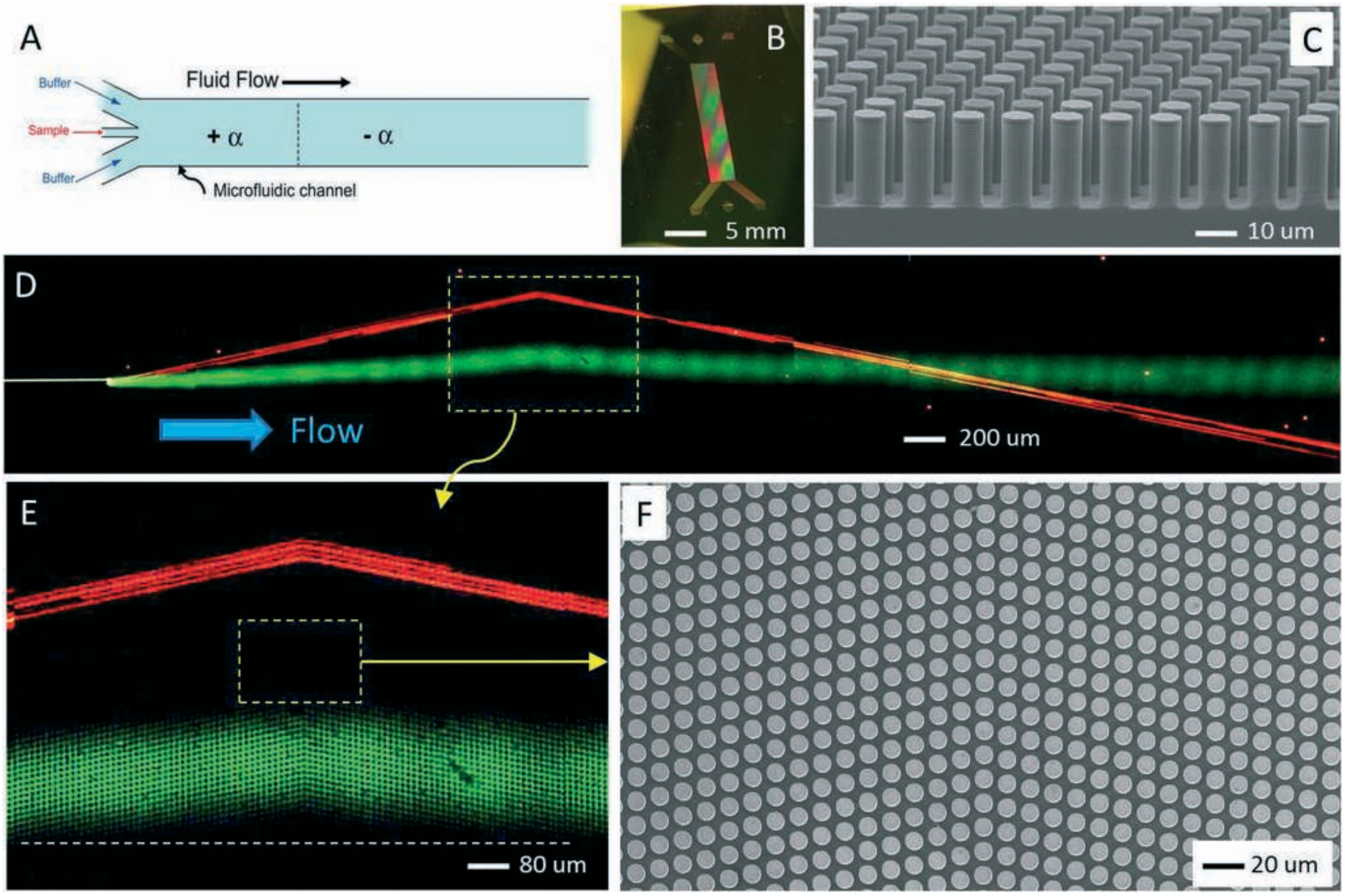

**Fig. 2** Device for demonstrating anisotropic flow tilt in the parallelogram layout. (A) Schematic of a DLD device layout having two sections with both positive and negative array inclinations. (B) Photo of the DLD device, as-etched into a silicon substrate. (C) Cross-sectional SEM image of the DLD array constructed from cylindrical posts on a 11 μm pitch parallelogram layout with an inclination of $\alpha = \pm 11.3°$ and a 3.7 μm lateral gap between posts. (D) Time exposure overlay showing the flow and separation of a mixed jet of fluorescent dye (green) and microbeads (2.7 μm, red) through the DLD array with the direction of the average flow indicated by the blue arrow. (E) Zoomed-in image showing flow deviation of the dye, near the interface between positive and negative array inclinations. The dashed (white) line indicates the horizontal direction. (F) Top-view SEM image showing the interface junction between the oppositely inclined array sections.

Fig. 3A captures the separation of 0.5 μm and 2.7 μm beads along the length of the device (Video S1†). In this case, 0.5 μm green beads which are nevertheless well below the critical particle size for bumping ($d_c$ = 2.4 μm) are also perturbed from the horizontal as they approach and leave the junction between the two array inclinations. Fig. 3B provides a zoomed-in view of the interface region (Video S2†). To further study the impact on bead trajectories, 1.9 μm green fluorescent beads were added to the input bead mixture. An intensified, monochrome CCD camera was used to capture the dynamic motion of beads with all three diameters as they track across the interface. Fig. 3C is a frame sum from that video which clearly shows a transition from zigzag to bump-type motion for 1.9 μm beads near the interface (Video S3†). The inset (Fig. 3D) is again a colour CCD exposure overlay following the paths taken by the particles with three sizes, well beyond the interface, at the cross-over point. Here, the distinct bump mode trajectories of the red 2.7 μm beads are seen alongside the brighter 1.9 μm green beads which have clearly reverted back to zigzag motion (a similar motion for the 0.5 μm bead stream is also just visible).

It is normally expected that all beads below the critical particle size for bumping should follow the zigzag path around array posts and transit the overall device horizontally. However, as we can see in Fig. 3, that while the beads initially track horizontally, the particles begin to mimic a bump trajectory as they approach the interface between the two sections, tracing the local array inclination (additional experimental evidence for anisotropic particle bumping is shown in Fig. S1†). Particles start to bump upwards at the end of the left array section, then immediately downwards at the beginning of the second array section. The particles then return to a horizontal trajectory as they continue into the middle of the second array. We suspect that these unusual and clearly undesired particle paths are the consequence of the lateral anisotropic flow acting on particle trajectories and that this behaviour stems from the inherent anisotropic permeability of the parallelogram layout. The average flow direction no longer remains horizontal and tilts towards the array inclination. As we shall see later with the help of simulations (sections 4.2 and 4.3), this flow tilt and the resulting reduction in the critical particle size become more pronounced near the interface between array sections. Such unintended particle behaviour in the zigzag mode is detrimental to separation, especially for high-resolution applications.[1,27,33,34]

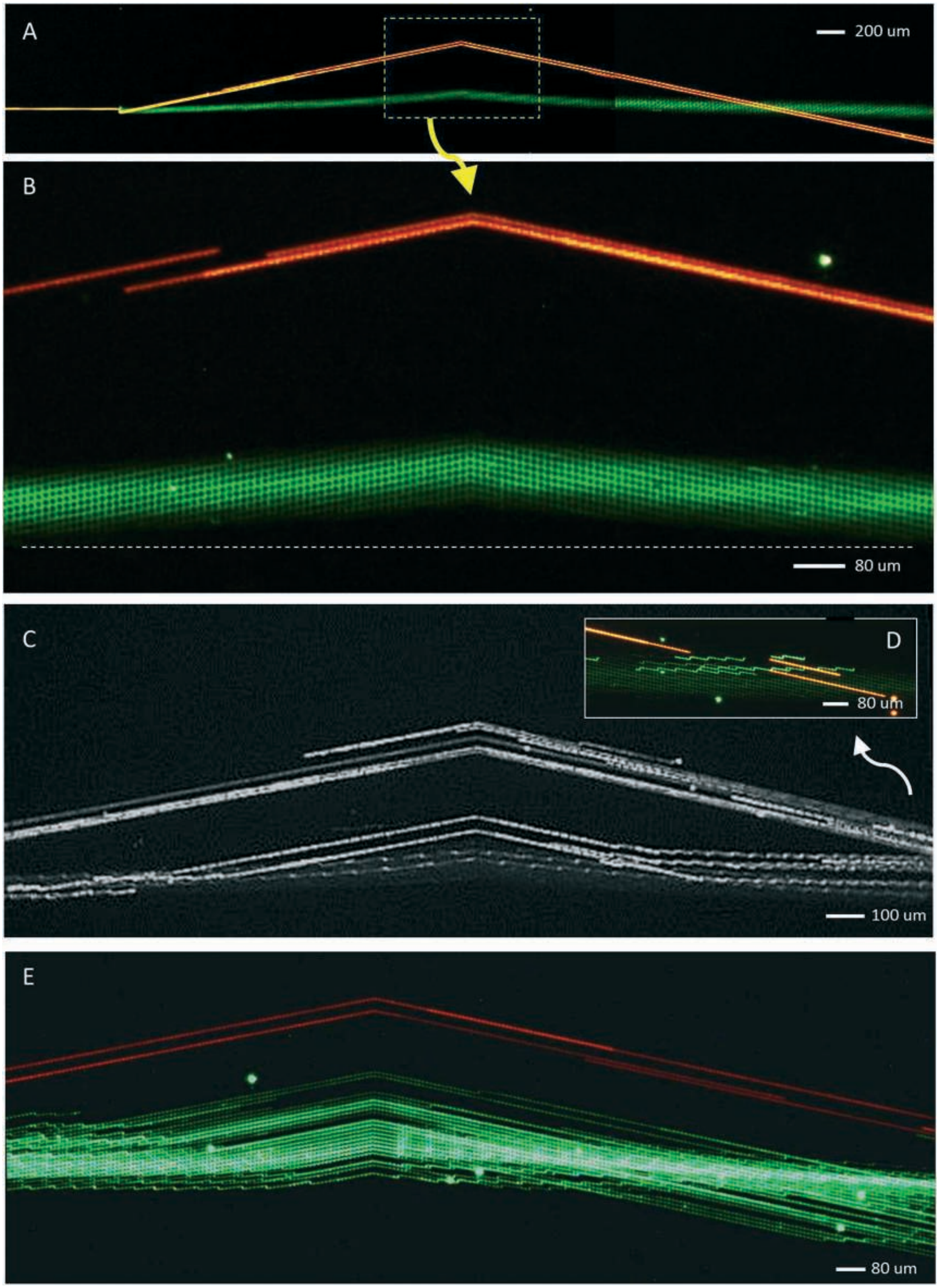

**Fig. 3** Device demonstrating the consequences of array anisotropy on bead trajectories. (A) Time exposure overlay (colour camera) showing a mixed jet of 0.5 μm (green) and 2.7 μm (red) microbeads injected into the same DLD device design as shown in Fig. 2 ($\alpha = \pm 11.3°$, lateral gap = 3.7 μm). (B) Zoomed-in image (colour camera, time exposure) showing the bump mode trajectory of the 2.7 μm beads and the path deviation for zigzag mode 0.5 μm beads near the interface between positive and negative arrays. The dashed (white) line indicates the horizontal direction. (C) Video frame sum (monochrome camera) of bead trajectories through the interface for a mixed bead jet that now includes 1.9 μm green fluorescent beads along with the 0.5 μm and 2.7 μm beads. Here 1.9 μm beads are expected to move in the "zigzag" mode through both sections ($d_c$ = 2.4 μm), but undergo anisotropic "bump" mode movement in the vicinity of the interface. Note that an intensified monochrome CCD camera was used to capture the dynamics of individual beads (SOM video 2). The inset (D, colour) details the crossover region beyond the interface in the negative inclination section. It shows that the 1.9 μm beads (brighter green) return to their expected zigzag mode downstream of the interface. (E) Time exposure (colour) detailing the addition of 2.3 μm beads (also green), which are close to the critical particle size for bumping. Notably, the 2.3 μm beads appear to be locked in the anisotropic "bump" mode well beyond the interface.

### 2.2 Rotated-square layout

In DLD devices with the rotated-square array layout, we do not observe significant flow deviation from the horizontal near the interface between two oppositely inclined array sections. The rotated-square DLD device shown in Fig. 4 has a positively inclined first section followed by a negatively inclined second section, following the form in Fig. 3. In this experiment, the array inclination is set to $\alpha = \pm 5.71°$ or $\varepsilon = 1/10$. The arrays are constructed on a denser 8 μm centre-to-centre pitch with a lateral gap of just 2.4 μm between 5.6 μm cylindrical posts ($d_c$ = 1 μm). In addition, sets of rounded rectangle structures were used here to match the adjacent sections of the rotated-square arrays (unlike parallelogram arrays which match intrinsically). Fig. 4A shows an integrated image of 0.5 μm and 1.9 μm green fluorescent beads moving across the mid-chip interface (Video S4†). At this point, the beads have already undergone lateral separation, following injection as a bead mixture into the DLD array as a narrow hydrodynamic jet. The 1.9 μm diameter beads are larger than the critical particle size and follow the canonical bump mode as expected, while the smaller, 0.5 μm beads follow zigzagging streamlines and act as fluid flow tracers.

Fig. 4B highlights individual particle paths of bump mode beads as they transit the interface. Note the switch between a bumping path along the upper side of the posts in the positively inclined array and that to the underside of the posts in the negatively inclined array. Fig. 4C similarly shows the averaged paths of 0.5 μm tracer beads as they cross the interface; the rotated-square geometry is clearly highlighted as the beads in the zigzag mode span all available streamline slots. The overall bead path for zigzag mode particles, which flows along the applied pressure gradient, does not deviate from the horizontal significantly (dashed line in Fig. 4A). This suggests the absence of anisotropy in the rotated-square layout and tolerance to the unintended particle trajectories observed for subcritical particle sizes in the parallelogram layout.

# 3 Nature of array-induced anisotropy

Anisotropic permeability[35] is the tendency of an array to induce a pressure gradient along the lateral axis (vertical in Fig. 1). When using an incompressible fluid, this pressure gradient is only problematic when it induces flow tilt along the lateral direction.[27,28] Fig. 2 and 3 demonstrate examples of such anisotropy effects, where the flow tilts along the array inclination. The varying flow tilt reduces the effective $\varepsilon$ locally and therefore also reduces the critical radius $r_c$. This change can cause unexpected particle bumping for particles expected to be in the zigzag mode, in the sections of the array. To avoid undesired spatially dependent $r_c$ and unintended particle trajectories, it is crucial to understand and control sources of anisotropy.

In 2007, James C. Sturm hypothesised that the parallelogram layout may display greater anisotropy than the rotated-square layout.[36] This hypothesis was drawn from the understanding of an optical phenomenon known as birefringence.[37] Optical birefringence, as observed in materials such as calcite, is caused by anisotropic optical transmission. In calcite, it is due to its non-cubic (parallelogram) unit cell. In contrast, optical materials with cubic unit cells show no anisotropic transmission and no birefringence. The analogy between optics and fluidics serves as a springboard for further investigation into array anisotropy in both of the array layouts that are usually treated as equivalent.

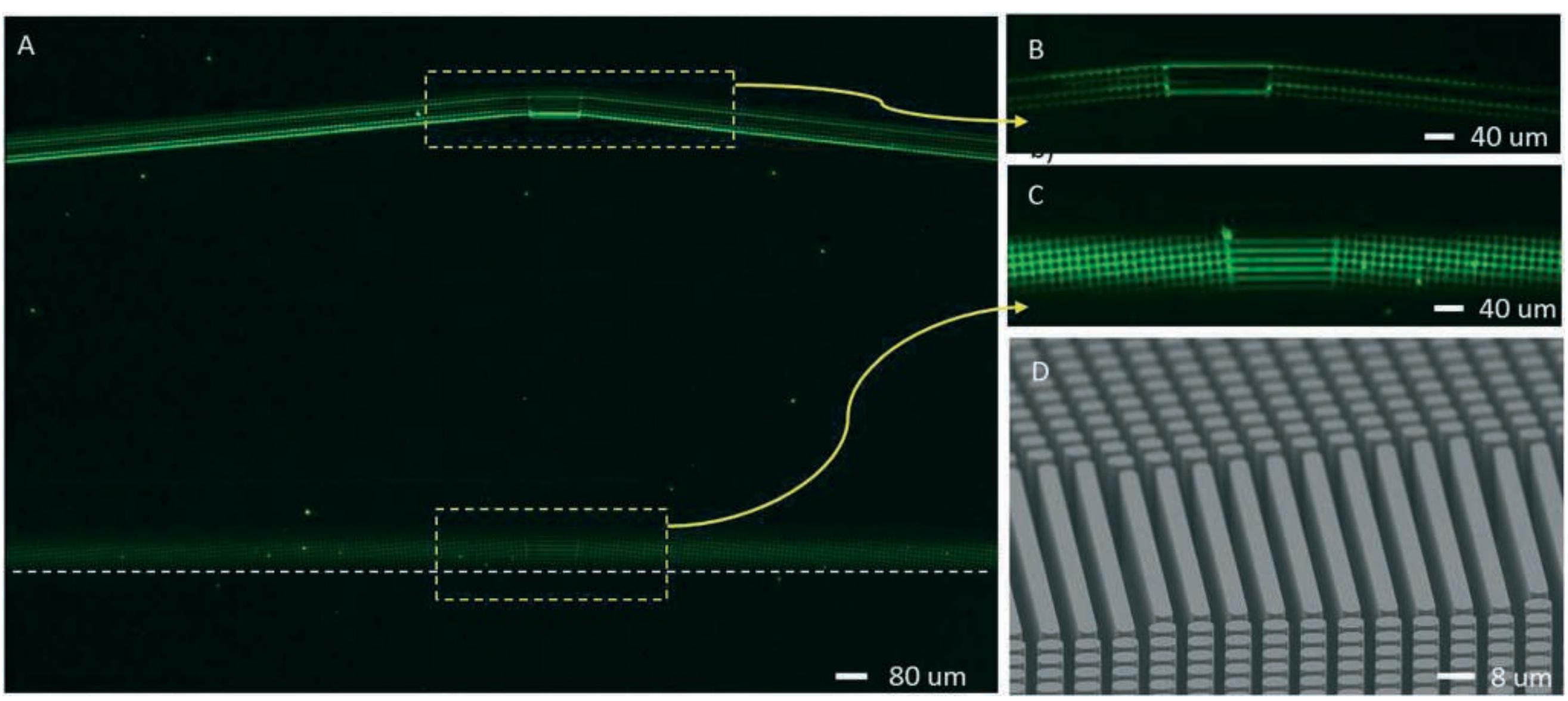

Fig. 4 Bead paths in rotated-square DLD arrays. (A) Time exposure (colour) of 1.9 μm (green) and 0.5 μm (green) fluorescent bead separation in a DLD array with the rotated-square layout. Similar to the parallelogram device in Fig. 3, this DLD has two adjacent array sections with positive and negative inclinations ($\alpha = \pm 5.71°$ or $\varepsilon = 1/10$). Both sections have $d_c$ = 2.4 μm (8 μm pitch, 2.4 μm lateral gaps). Insets (B and C, colour) detail bead trajectories across the interface region for both the 1.9 μm bump mode beads as they cross the interface and the overall path of the 0.5 μm zigzagging beads. For the tracer beads in the zigzag mode, no significant deviation from the horizontal (dashed white line) is observed. (D) A tilted SEM image of the interface structure used here to match the two adjacent rotated-square arrays.

# 4 Mapping anisotropic permeability

We put the hypothesis that the parallelogram layout possesses greater anisotropic permeability than the rotated-square layout to the test by using high-resolution lattice-Boltzmann simulations in the Stokes flow limit. Simulations are run in two dimensions and carried out over a single post of the array domain (400 × 400 lattice cells) under periodic boundary conditions (section 7.1). This approach simulates flow over a central post of an infinite obstacle array. We also carry out large-domain full-array simulations, with as many as 152 × 120 posts in the device in order to investigate the flow tilt due to array anisotropy.

### 4.1 Anisotropic lateral pressure drop

For the purpose of generality, we begin with a symmetric circular post shape with a gap-to-post diameter ratio ($G/D$) of unity. The simulated device gap and post diameter are 10 μm each (post–pitch distance, $\lambda$ = 20 μm). Simulations are carried out over the entire range of the row shift fraction ($\varepsilon$ = 0.0 to 1.0, at 0.1 increments) for both the parallelogram and rotated-square layouts.

Using periodic boundary conditions to simulate a unit cell (*i.e.* a single post) of an array possessing anisotropic permeability would result in non-zero average lateral fluid velocity. In real microfluidic devices, this lateral flow is restricted by microchannel side walls. Therefore, we quantify array anisotropy by measuring the lateral pressure drop that is required to maintain the zero average lateral fluid flow. We define anisotropy as the dimensionless ratio of the induced lateral pressure gradient to the imposed pressure gradient along the flow direction (Fig. 5A). Fig. 5B shows anisotropy values mapped for various inclinations ($\varepsilon$) for both the parallelogram and rotated-square layouts. The sign of the anisotropy value indicates the direction of the lateral pressure drop; a positive sign means that the lateral pressure drop is in the same direction as the row shift.

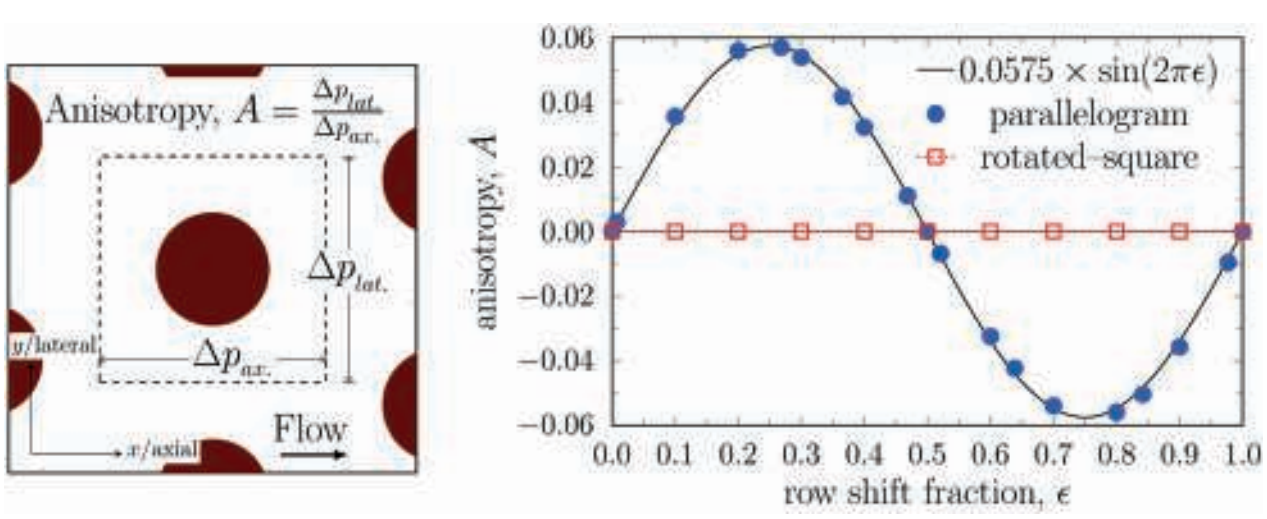

Fig. 5 The left panel illustrates the definition of array anisotropy in our simulations. $\Delta p$ is the pressure drop over one unit distance for the array. The right panel shows the anisotropy variation in parallelogram and rotated-square layouts (circular posts) for changing array inclination, $\varepsilon$.

For the parallelogram array, the anisotropy shows sinusoidal dependence on $\varepsilon$. Moreover, the absolute anisotropy values are equal for $\varepsilon$ and $1 - \varepsilon$. This follows from the fact that a parallelogram array with $0.5 < \varepsilon < 1.0$ is equivalent to one with $1 - \varepsilon$, but with a negative row shift. We observe a maximum anisotropy of ≈5.6% occurring at $\varepsilon$ = 0.25 for the parallelogram array.

The rotated-square layout, however, exhibits vanishing anisotropy for all tested values of $\varepsilon$. This corroborates the hypothesis of the rotated-square layout having an advantage over the parallelogram layout in avoiding anisotropic effects.

### 4.2 Anisotropic flow tilt

Array anisotropy can only affect particle trajectories when it causes a tilt in the flow direction. For the parallelogram layout with symmetric circular posts, the direction of anisotropy is the same as the row shift. Therefore, the anisotropic flow tilt occurs towards the array incline and causes a decrease in the effective array inclination.

To demonstrate the reduction of the effective inclination, we simulated the mid-section of two DLD devices (Fig. 6), one with the parallelogram layout (Fig. 6B) and the other with the rotated-square layout (Fig. 6C). Each device has 152 circular posts along the flow and 120 posts along the transverse direction. The post diameter is 10 μm, and the gap between posts is 10 μm. The simulated domain is 3.2 mm × 2.4 mm (3200 × 2400 lattice cells) with periodic inlet and outlet flow conditions (Fig. 6A). The flow is driven by a pressure gradient along the axial direction. Each device has two array sections with opposite inclinations. The left section has a positive array inclination of $\varepsilon$ = 0.2 and the right section has an inclination of $\varepsilon$ = −0.2. Both sections are separated by a gap of ≈4 posts (80 μm).

Flow streamlines (blue lines) are shown (Fig. 6B and C) for both devices. In the parallelogram device (Fig. 6B), the streamline nearest to the right side wall remains horizontal throughout. As they move away from the right side wall toward the centreline of the device, the streamlines start tilting along the prevalent array inclination. Already ten posts away from the right side wall, this effect becomes important and the tilt continues to increase as they move further away from the side wall. In the central region of the device (typically the particle separation zone), the flow is no longer parallel to the side walls of the device. We also observe a similar behaviour at the left side wall (data not shown).

Around the centreline, the effective array inclination is reduced from 1/5 to ≈1/7. This change in $\varepsilon$ occurs gradually with position and is therefore rarely equal to one over an integer. It is known that such non-integer periodicity values for bump arrays can cause multi-directional sorting modes as well as negative directional locking.[21,25] All of these effects are highly undesirable for deterministic bimodal particle sorting. Furthermore, secondary factors, such as the distance from the side walls, now influence the critical radius.

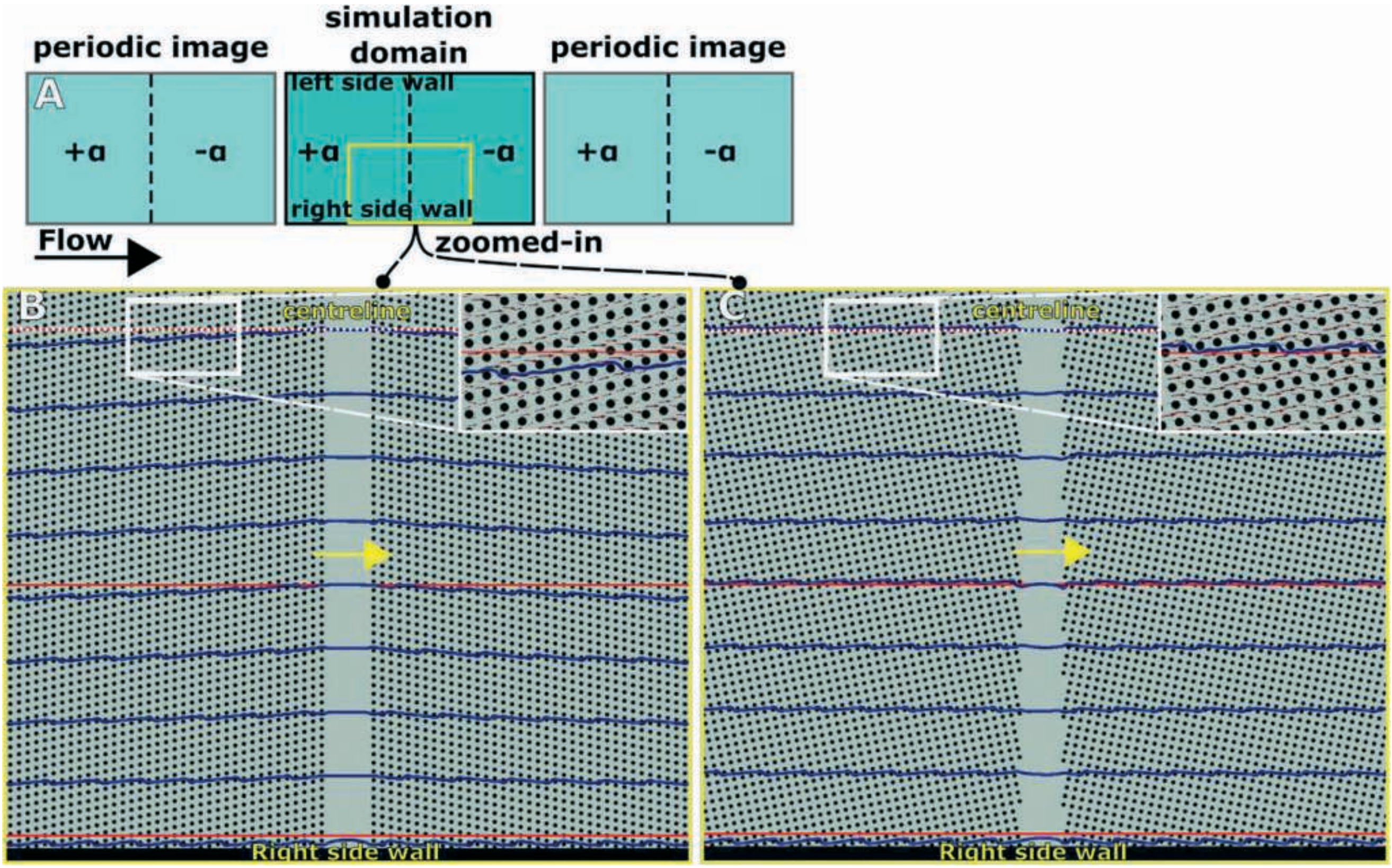

**Fig. 6** The simulation setup for both parallelogram and rotated-square DLD arrays is shown in (A). The device has two counter-inclined array sections with $\varepsilon$ = 0.2 and periodic inlet and outlet boundaries. Fluid streamlines are plotted (blue) in (B) the parallelogram and (C) rotated-square layout devices. Reference horizontal lines (red) indicate the direction parallel to the applied pressure drop (parallel to side walls). (B) Away from the right side wall, the streamlines tilt significantly in the parallelogram layout device. (C) The streamlines remain parallel to the applied pressure drop and follow the "zigzag" path in the rotated-square layout device. The small deviations from the horizontal line in the latter case are due to the streamlines navigating around posts. The arrows in the zoomed-in insets indicate the local velocity field. The device centreline is shown as the dashed white line.

In the rotated-square device (Fig. 6C), the streamlines (blue lines) are horizontal. The streamlines remain on their predicted course (horizontal "zigzag" around the posts), parallel to the side walls of the device. This can be attributed to the absence of anisotropy. Unlike the parallelogram layout, using the rotated-square array leads to a well-defined and constant critical radius throughout the device. Therefore, the rotated-square layout should be preferred for particle separation applications.

### 4.3 Background secondary recirculation

We have seen that the inherent anisotropic permeability of the parallelogram layout can tilt the streamlines away from the side walls, along the prevalent array inclination. However, this need not always be the case. Under certain conditions, the array anisotropy leads to a lateral pressure gradient that is balanced by normal stresses at the side walls. In this case, the streamlines are not tilted and particle trajectories are not affected. As we shall see later, certain common DLD design features, however, allow the lateral pressure gradient to induce secondary flows that tilt the streamlines.

By investigating the flow field in the parallelogram device (Fig. 6B), we find that the anisotropic lateral pressure drop begins to be released near the array section interface manifesting as secondary recirculation in the device. We plot this complex "ladder-like" background flow recirculation pattern in Fig. 7. This secondary flow field is obtained by subtracting the axial velocity component at the centre of the device ((at 1.6 mm, 1.2 mm) marked as X) from the overall primary velocity field. The circulation is clockwise in this case, and meanders around the posts in the array. The recirculatory flow causes the streamline tilt which in turn alters the critical radius locally. We find that the recirculatory flow is absent when the rotated-square layout is used. As demonstrated next, secondary recirculatory flow manifests when certain device design features or "enablers" are present in devices with intrinsic anisotropy.

### 4.4 Anisotropic flow tilt "enablers"

Certain design features that allow the anisotropic pressure gradient to drive the recirculatory flow are quite common in DLD devices. In general, placing device sections with a large difference in their anisotropy values next to one another allows the background flow to develop, such as when arrays with opposite anisotropy values are placed next to each other (as shown in Fig. 2, 3 and 6). The anisotropic flow tilt is also "enabled" when sections with a significant difference in anisotropy values or a low-impedance isotropic zone, such as an interface gap between array sections, are employed in the DLD design.

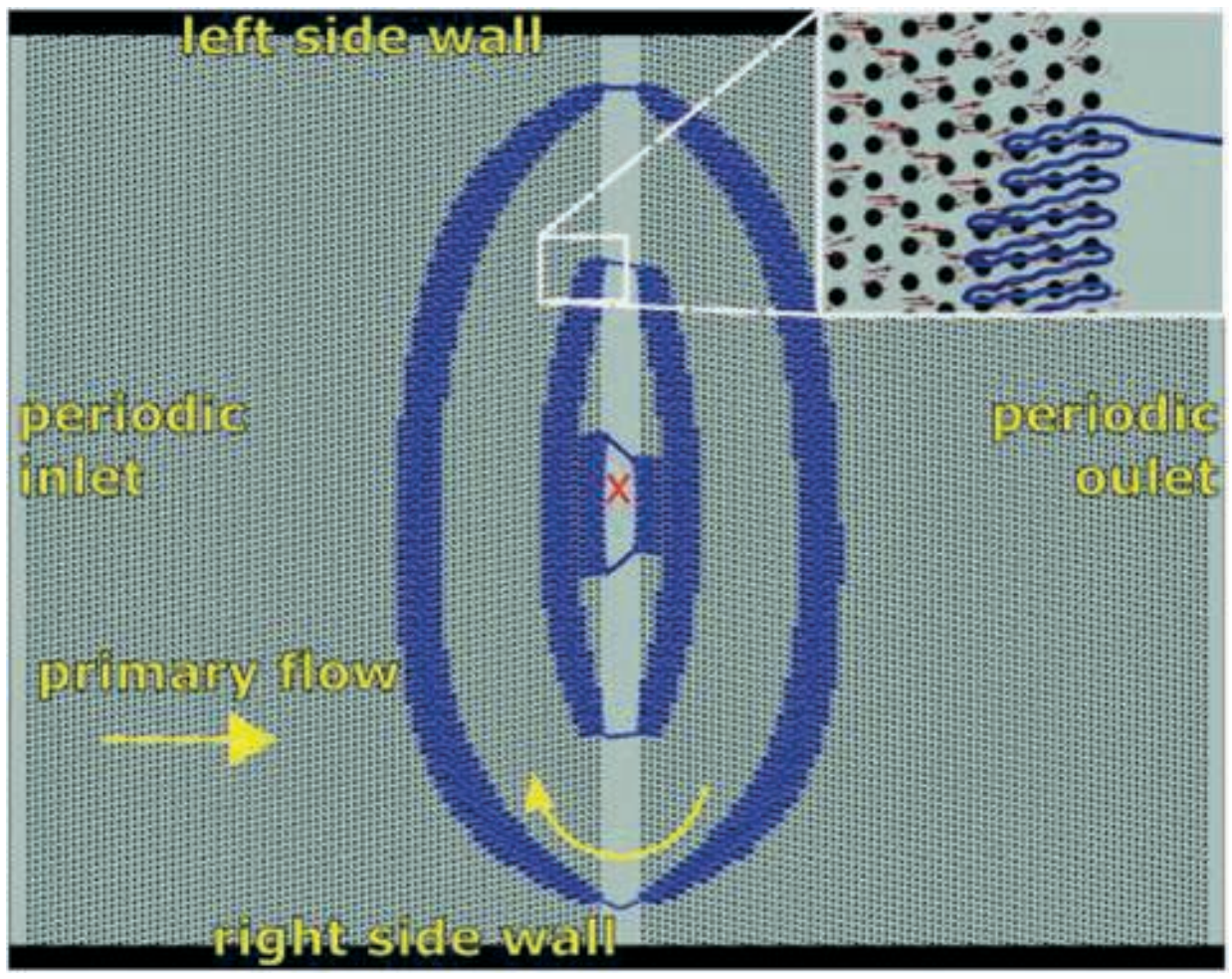

**Fig. 7** Secondary recirculation flow pattern streamlines (blue) in the parallelogram device with $\varepsilon$ = 0.2 (Fig. 6A) obtained by subtracting the $x$-component of the primary velocity measured at the device centre (marked: X) from the overall velocity field. The arrows in the zoomed-in inset indicate the local secondary recirculation velocity field.

To demonstrate the effect of an interface gap between device sections, we carried out two simulations with the anisotropic parallelogram array with $\varepsilon$ = 0.25. One device features an interface gap between sections (Fig. 8A), while the other does not (Fig. 8B). In the device in Fig. 8A, the interface section gap acts as an enabler by allowing the fluid flux to compensate for its upward anisotropic tilt in the arrays. This allows the flow to tilt along the prevalent array anisotropy. In the absence of the gap, the flow remains horizontal everywhere (Fig. 8B). Additional simulations show that providing connector elements in the interface gap (such as those in Fig. 4D) also suppresses lateral flow tilt. No flow tilt is observed when the gap between these connector elements is kept equal to the array gap $G$. Here, the connector elements mimic the no gap situation by preventing fluid flux deviation from the horizontal. However, when the spacing between these connector elements is increased ($>4\lambda$), the flow tilt is seen to gradually manifest again.

Even without an interface gap, the anisotropic flow tilt manifests when two array sections with significant differences in the anisotropy magnitude or direction are used next to one another (cascaded array). We carried out a simulation of a cascaded parallelogram layout device with the left array section at $\varepsilon$ = 0.05 and the right array section at $\varepsilon$ = 0.25, with no interface gap in between (Fig. 8C). The array section with higher anisotropy dominates and causes complementary flow tilt in its adjacent array sections. Fig. 8C shows that, away from the side walls, the flow tilts slightly upwards in the right array section ($\varepsilon$ = 0.25) and, in order to compensate for this tilt, slightly downwards in the left array ($\varepsilon$ = 0.05). Here, the effective array inclinations (in the central simulation zone) become $\varepsilon$ = 0.231 and $\varepsilon$ = 0.068 in the right and left sections, respectively. Therefore, we find that the cascaded parallelogram arrays may generally have a locally varying critical radius $r_c$.

In our simulations, we observe that the anisotropic flow tilt occurs at the entrance and exit regions of similarly-inclined array sections, when an interface gap ($<6\lambda$) is present. The length of the array region affected by anisotropic flow tilt scales with the strength of anisotropy and the width of the array. We find that the length of the anisotropy-affected zone protruding into the array does not exceed the array width. Therefore, a zone free from the anisotropic lateral tilt can be obtained at the array centre by fabricating arrays with more than twice the number of posts in the flow-wise direction compared to those along the lateral direction. We suggest such "long and narrow" array sections in order to have a sufficient number of pillars in the centre, unaffected by the anisotropic flow tilt and therefore having invariant $r_c$, for the particles to separate.

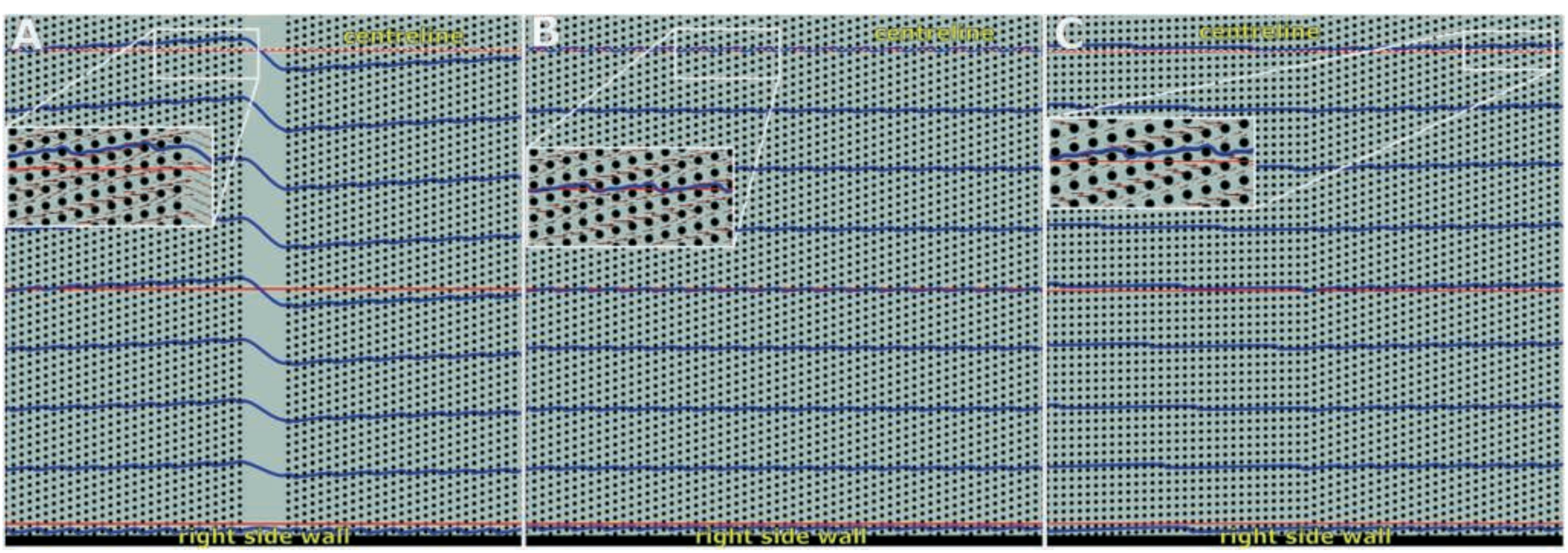

**Fig. 8** Fluid streamlines (blue) in a device (A) with an interface gap between array sections and (B) without an interface gap. The devices in (A) and (B) have parallelogram layout with circular posts and $\varepsilon$ = 0.25 in both sections. There is a significant flow tilt in (A) and no tilt in (B). (C) Streamlines in a cascaded DLD with $\varepsilon$ = 0.05 (left section) and $\varepsilon$ = 0.25 (right section). Horizontal lines (red) indicate the axial direction with the device centreline indicated (dashed white line). All panels are zoomed-in views of larger DLD devices, taken near the right side wall. Arrows in the insets indicate the local velocity field.

An interface gap is often seen at the beginning and end of arrays in most DLD devices. Such gaps should be avoided. The cascade arrangement of arrays is commonly employed for separation of more than two particle species in a single device. In such applications, the rotated-square layout should be used in cascade arrangement, rather than the parallelogram layout. It is often the practice to suppress all lateral flow in simulations for the design of DLD devices.

In reality however, we see that the side walls do not confine the lateral flow everywhere in the presence of these "enablers". Therefore, as a general rule, "enabler" features that allow the flow to locally recompense for the lateral flow tilt should be avoided and the rotated-square layout should be favoured.

## 5 Causes of excessive anisotropy

Non-circular post shapes, non-unity array aspect ratios and post-to-gap ratios have been used widely in the DLD in recent years. Here, we focus our attention on the inherent anisotropy of such non-standard arrays which would give rise to lateral flow tilt in the presence of the "enablers" discussed in the previous section. In this section, we carry out single post simulations to map array anisotropy (as lateral pressure drop) as well as large domain simulations to visualise the anisotropic flow tilt.

### 5.1 Unequal axial-to-lateral post distance

DLD devices with unequal axial and lateral gaps between posts have been shown to give enhanced separation in specific applications.[5,30] Using such non-unity aspect ratios for the array unit cells clearly has value, but we show here that there is a cost in terms of higher anisotropy. As previously discussed, we carried out single post simulations to study the effect of the post–pitch aspect ratio on array anisotropy. The aspect ratio is quantified as AR = $\lambda_a/\lambda_b$ (Fig. 1A and B). Here, we vary the axial gap $\lambda_a$ and the lateral gap and pillar diameter are both kept equal to $G = D = \lambda_b/2$. All other simulation parameters are the same as before.

Fig. 9A shows the variation of anisotropy at array inclinations of $\varepsilon = 0.1$, 0.3, and 0.5 for the parallelogram array. Inclinations of $\varepsilon > 0.5$ are equivalent to a negatively inclined array with inclination $1 - \varepsilon$ and are not plotted. For $\varepsilon = 0.5$, the anisotropy must vanish for all aspect ratios due to symmetry reasons.

Interestingly, in the parallelogram array, the anisotropy steadily decreases and converges to zero with increasing AR or $\lambda_a$. This is an important result for reducing the anisotropic permeability in parallelogram arrays, especially since the critical radius $r_c$ is independent of the aspect ratio, at constant $\varepsilon$ and $G$ (our simulations predict $r_c$ to be 1.8 μm for $\varepsilon = 0.1$ and 3.6 μm for $\varepsilon = 0.3$, independent of the aspect ratio). However, AR > 1 has a clear disadvantage; large aspect ratios mean longer devices for the same lateral displacement. This raises issues of greater device footprints and higher fluidic resistance. Therefore such arrays are normally not used in practice.

Fig. 9B shows the anisotropy values for inclinations $\varepsilon = 0.1$, 0.3, and 0.5 for the rotated-square array. We observe that the sign of the anisotropy changes when the aspect ratio crosses the value 1. For AR < 1, the anisotropy is positive and for AR > 1, it is negative. AR = 1 leads to zero anisotropy for all investigated values of $\varepsilon$. Therefore unless other requirements call for non-unity aspect ratios in rotated-square arrays, AR = 1 should be chosen. If the aspect ratio is not unity, the anisotropy can be reduced by decreasing $\varepsilon$.

### 5.2 Unequal array gap-to-post size ratio

Unequal array gap to post-sizes ($G/D \neq 1$) are very common in DLD arrays. Fig. 9C shows the variation of anisotropy with the change in this ratio for the parallelogram array. Here, as the size of the post relative to the array gap increases, so does the anisotropy. We can see that the highest anisotropy value is significantly lower than that induced because of the changes in the array aspect ratio. These results indicate that using larger gap sizes relative to the post-size would reduce anisotropy in the parallelogram array. Qualitative experimental evidence for reduction in the anisotropic flow tilt with increased array gap-to-post size ratio is shown in Fig. S2.† We find that in the rotated-square layout, the gap-to-post size ratio does not have any effect and anisotropy remains zero.

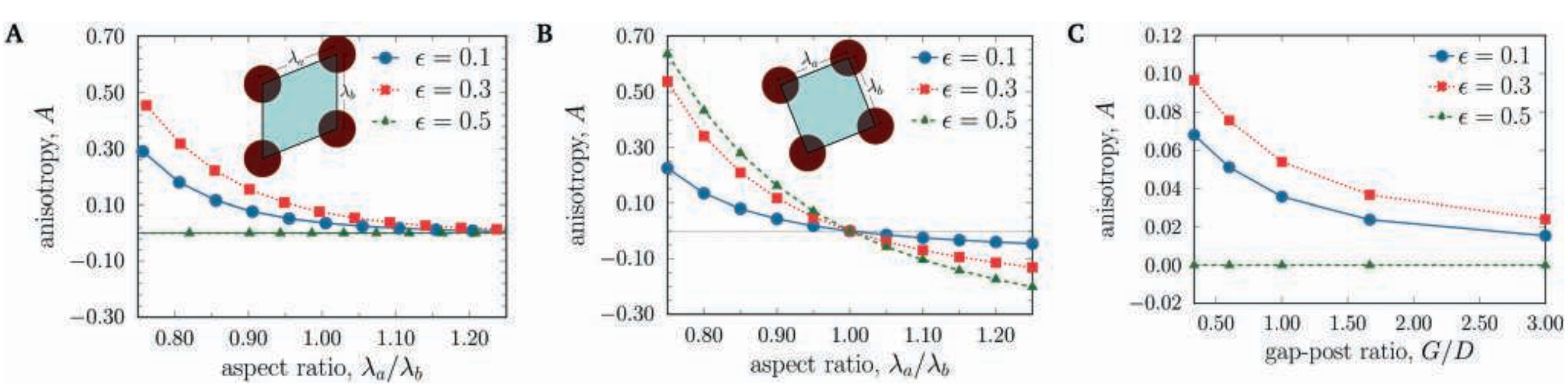

**Fig. 9** Variation of anisotropy with the aspect ratio of the array unit cell in (A) the parallelogram and (B) the rotated-square layouts. The corresponding unit cell layouts are indicated in the insets. (C) Variation of anisotropy with the change in the gap to post-diameter ratio of the array for the parallelogram layout.

### 5.3 Post shape-induced anisotropy

We find that asymmetric post shapes can cause a severe increase in anisotropy. Fig. 10 displays images from an experiment with right-triangular posts arranged in the rotated-square layout with negative array inclination ($\varepsilon$ = −0.1). Fig. 10A shows the interface gap between the two array sections. The cylindrical pillars placed in the gap are roof supports. All array parameters are equal in both the left and right sections (6 μm post size, 4 μm gap, row shift fraction $\varepsilon$ = −1/10). However, the right-triangular posts are rotated by 90° counter-clockwise in the right section relative to those in the left section.

Fig. 10B shows trajectories of fluorescent beads with 3.1 μm diameter. The beads are larger than the design critical radius ($r_c$ = 1.1 μm on the vertex side and $r_c$ = 1.5 μm on the flat side of the triangle). Therefore, the beads are expected to follow the bump trajectory moving downwards along the array inclination. Instead, we can see that the beads move along an abnormal "zigzag" trajectory in the left array section. However, in the right array section, the same beads start following the "bump" trajectory. In the right section, close to the interface gap, the beads bump on the flat side of the right triangular posts, rather than on the vertex side as was intended. This unexpected behaviour is due to the anisotropic flow induced in the device caused by the strongly anisotropic triangular post shape. The flow pattern in both sections tilts along the hypotenuse of the triangle (rather than the array incline), thereby increasing the effective negative inclination in the left section and decreasing it in the right section. In fact, particles bumping on the flat side of the triangles in the right section indicate that the flow tilts beyond the array incline $\alpha = \tan^{-1}(\varepsilon) = -5.7°$, effectively creating a positively inclined array region close to the central interface gap.

To visualise the streamline tilt, a fluorescent dye was introduced into the bottom section of the DLD device (Fig. 10C). The local deviation of the flow is marked out by the interface between the dye and non-dye regions. This clearly reveals that the flow inclination is no longer horizontal and aligns with the hypotenuse of the triangular posts in the array segments. We see that, away from the side walls, the flow deviates by as much as ≈250 μm from the horizontal. Such a large deviation arising from the anisotropic pillar shape therefore induces completely opposite particle behaviour to that intended.

To corroborate this claim, we simulated a device mimicking the experiment with 160 × 120 triangular posts along the flow and transverse directions, respectively. The inclination is $\varepsilon = -0.1$ and the device parameters match those in the experiment (6 μm post size, 4 μm gap). The boundary conditions for the simulated device (3200 × 2400 lattice cells) are the same as those in Fig. 6A. The simulation domain has two array sections, with the triangular posts in the right section rotated by 90° relative to those in the left section. The central interface gap is ≈50 μm and simulated without the cylindrical roof supports.

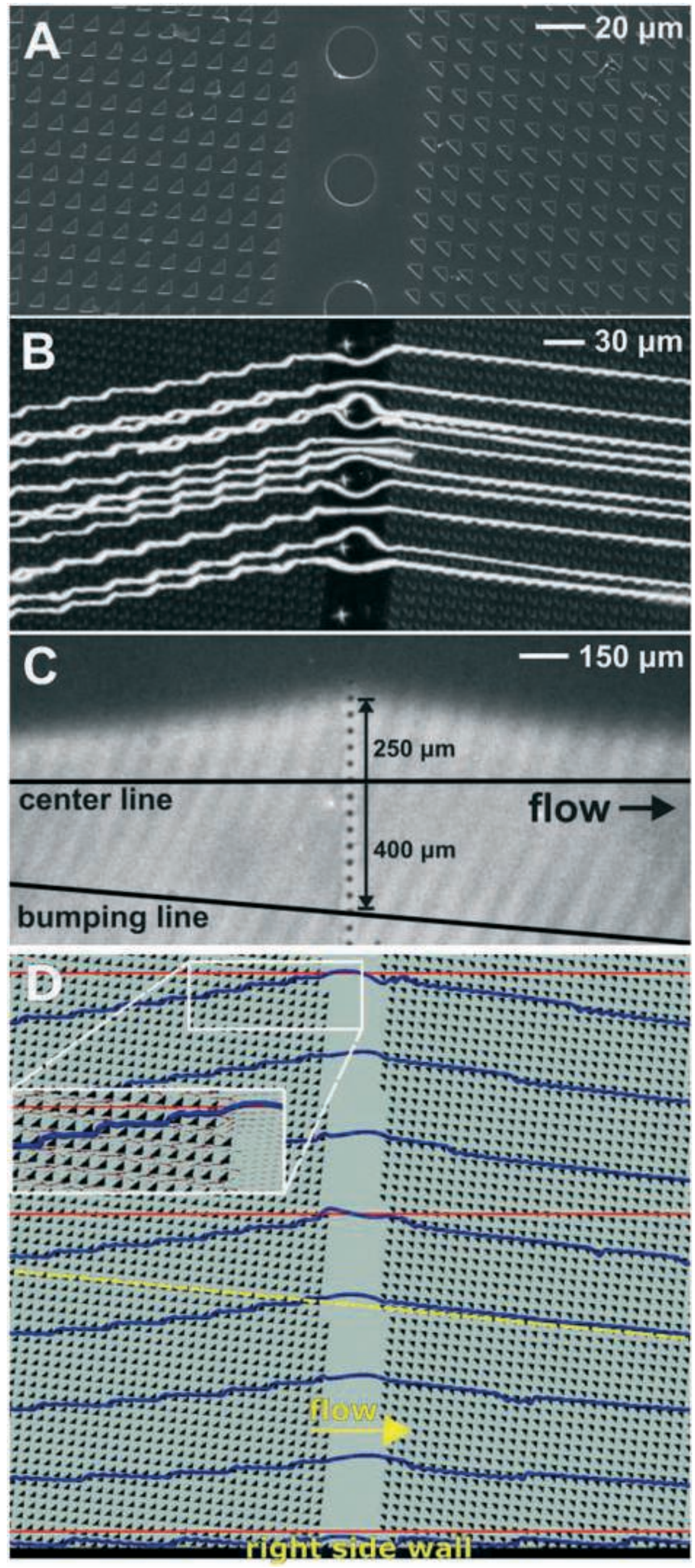

Fig. 10 Experimental device using right-triangular posts with the rotated-square layout. (A) Scanning electron micrograph of the junction between two array sections. The array inclinations are identical on either side ($\varepsilon$ = −0.1), but the triangular posts are rotated by 90° counter-clockwise. (B) Epifluorescence micrograph showing trajectories (white lines) of fluorescent beads with a super-critical diameter of 3.1 μm. In the left section, the beads are in the zigzag mode, while in the right section, they travel in the bump mode. (C) Path of fluorescent dye injected along the right side wall of the device as it crosses the interface junction. The array inclination, centre line and maximum deviation of the dye are marked. (D) Flow streamlines (blue) from a simulation of the triangular DLD geometry. Significant tilt in the streamlines from the horizontal (red lines) is observed away from the device side walls. The dashed yellow line indicates the negative array inclination. Arrows in the inset indicate the local velocity field.

Fig. 10D shows a subset of the simulation domain at the right side wall with the flow streamlines (blue). The horizontal (red) lines indicate the direction of the applied pressure gradient. Around the tenth post from the right side wall, we can see that the flow tilts significantly along the triangle post hypotenuse. The tilt increases to a maximum as we move towards the centre of the device. We approximate the flow tilt in the central zone, near the central interface gap, by averaging over six equi-spaced streamlines. For the left array section (the triangles pointing up), the flow tilts by ≈ +11.3°. This would correspond to an effective array inclination of $\varepsilon \approx -0.31$ and an empirically estimated[2] critical diameter of ≈3.19 μm. In the right array section (triangular posts pointing down), we measure the tilt to be ≈ −11.9° and therefore an effective array inclination of $\varepsilon \approx +0.11$ and an estimated critical diameter of ≈1.94 μm. These results support the experimental observations of having beads, 3.1 μm in diameter, in the "zigzag" mode in the left section and in the "bump" mode in the right section.

**Anisotropy of non-circular posts.** To understand the role of the post shape better, we simulated a single post with periodic boundary conditions (400 × 400 lattice cells) and measured the anisotropy for different post shapes commonly employed in DLD devices. We tested the square,[16] equilateral triangle,[12,19] right triangle[18] and I-shape[16,17] posts in a rotated-square layout for an inclination of $\varepsilon = 0.1$. All the posts are defined such that they can be inscribed in a circle of 10 μm in diameter. Each post is rotated to align with the array inclination at $\varepsilon = 0.1$. The results are collected in Table 1, along with those for the circular post (10 μm diameter). We see that highly asymmetric post shapes, such as the right triangle, display anisotropy an order of magnitude higher than the maximum due to the parallelogram layout with circular posts. However, the anisotropy of other post shapes is close to zero and lower than that of the parallelogram layout with cylindrical posts. Therefore, the anisotropy caused by the device layout can be more important than the post-induced anisotropy, and the rotated-square layout is generally preferable.

Table 1 Anisotropy for different post shapes in a rotated-square layout with an inclination of $\varepsilon = 0.1$. The finite anisotropy for the circular post ($O(10^{-7})$) is caused by numerical approximations. Note that anisotropy for the parallelogram layout array with circular posts at $\varepsilon = 0.1$ is $3.6 \times 10^{-2}$

| Post shape | | Anisotropy, A |
|---|---|---|
| Circular | | $3.1 \times 10^{-7}$ |
| Square | | $2.1 \times 10^{-6}$ |
| Equilateral triangle | | $3.2 \times 10^{-3}$ |
| Right triangle | | $1.8 \times 10^{-1}$ |
| I-shape | | $5.2 \times 10^{-3}$ |

A highly anisotropic asymmetric post shape such as the right triangle can prove useful. In case an anisotropic array needs to be employed, such a post shape can be used in order to cancel out array anisotropy. The anisotropy of post shapes can be varied by rotating them with respect to the flow direction. We plot the anisotropy variation for the post shapes listed in Table 1, for different degrees of rotation with respect to the flow direction in Fig. S3.† The simulations are carried out for each post shape at a given angle with respect to the flow in a rotated-square layout with an inclination of $\varepsilon = 0.1$. These post shapes have the same size as those discussed earlier. Thus, anisotropic post shape rotation could be used in order to obtain zero net anisotropy for any DLD array.

# 6 Suppressing the anisotropic lateral flow

We have demonstrated the existence of anisotropic flow tilt in the parallelogram array as well as when non-cylindrical asymmetric post shapes are used in the DLD. We see that the flow tilt can cause a mixed mode for the particle due to locally varying critical diameter in the array. Avoiding the problems associated with the anisotropic flow tilt is important for predictable separation of particles in the DLD. We give the following design points to the DLD user community to suppress the lateral flow tilt seen with the DLD. These are informed by both the simulations and experiments presented in this work.

• The rotated-square layout with cylindrical posts should be preferred over the parallelogram layout.

• "Enablers" such as interface gaps and counter-inclined adjacent sections should be avoided with anisotropic arrays.

• Increase the flow-wise to lateral array gap ratio to decrease anisotropy in the parallelogram layout.

• Anisotropic post shape rotation can be used to counter array anisotropy when using non-cylindrical posts.

• Use "long and narrow" arrays with a greater value of the post ratio in the flow-wise to lateral direction ($\gg 2$) in order to provide an adequate region possessing a constant critical radius $r_c$.

# 7 Materials and method

## 7.1 Simulation details

The simulations were carried out using our validated lattice-Boltzmann code.[29] The no-slip wall boundary condition is implemented using the standard half-way bounce-back model. The relaxation time is set to unity with the standard

BGK collision operator. All the simulations are in the Stokes flow regime with a Reynolds number (computed based on the maximum velocity and array gap, $G$) of $\mathrm{Re} < 0.8$ for the single post simulations and $\mathrm{Re} < 1 \times 10^{-4}$ for the large domain simulations.

### 7.2 Experimental procedure

The microfluidic devices used in this work were fabricated by standard photolithographic techniques and deep reactive ion etching as described in ref. 22. Approximately 1 mm diameter through-holes were sand-blasted using a dental sandblaster. The devices were sealed using a large PDMS coated glass coverslip and wet by immersion in water containing 2 g $L^{-1}$ Pluronic F108 (BASF), as detailed in ref. 38. Fluorescent polystyrene beads were diluted in ultrapure water containing 2 g $L^{-1}$ F108 and thoroughly sonicated to break up aggregates prior to being introduced into the devices.

## 8 Conclusion

We investigated anisotropic permeability in deterministic lateral displacement (DLD) arrays *via* experiments and lattice-Boltzmann simulations. Anisotropic devices induce a pressure gradient perpendicular to the axial flow direction. DLD anisotropy can manifest as undesired localized secondary flows (*e.g.* recirculation patterns). Secondary flows are undesired as they cause the imposed flow to tilt away from its intended axial direction, which in turn leads to locally varying critical separation size and unintended particle trajectories.

We found that the parallelogram layout displays inherent anisotropy that increases with increasing array inclination with respect to the axial flow ($0 < \varepsilon \leq 0.25$). In contrast, the rotated-square layout with circular posts shows no anisotropy and therefore no flow tilt. Hence, in the rotated-square array, the flow remains parallel to the side walls throughout, and the critical radius is predictable. We thus recommend the rotated-square layout, rather than the parallelogram layout.

Also, unequal axial and lateral post gaps and non-circular post shapes can lead to excessive array anisotropy, even for the rotated-square layout. While square, equilateral triangle and I-shaped posts lead to relatively low anisotropy, right triangle posts cause large anisotropy that can lead to significant streamline tilt. Increasing the array post-gap ratio also leads to increased anisotropy in the parallelogram array.

If anisotropic arrays are used, one should avoid "enabler" design features that allow the anisotropy to trigger off-axis lateral flows. One typical enabler feature is the interface gap between array segments in cascaded DLD devices or at the beginning or end of arrays.

Anisotropic permeability plays an important role in determining the success or failure of a DLD device and needs to be accounted for when designing such separation arrays. Additionally, this study of anisotropic permeability is relevant to a large class of flows in microfluidics and porous media, where the fluid has to flow past an ordered periodic set of obstacles, akin to those in the DLD.

## Conflicts of interest

There are no conflicts to declare.

## Acknowledgements

R. Vernekar thanks Dr. P. Valluri for access to the ECDF computational facilities at Edinburgh. T. Krüger thanks the University of Edinburgh for the award of a Chancellor's Fellowship and part-funding for this work. D. Inglis acknowledges support from ARC DP170103704 and thanks James C. Sturm for suggesting the optical birefringence analogy, the inspiration for this work, in 2007. The authors thank one of the reviewers for a detailed and constructive review.